\documentclass[10pt,final,journal,letterpaper,twoside,twocolumn]{IEEEtran}
\usepackage{graphicx,cite,epsfig,amssymb,amsmath,subfigure,float}

\begin{document}
\markboth{IEEE TRANS. ON VEHICULAR TECHNOLOGY, Vol. XX,
No. Y, Month 2014} {Ge etc.: Energy Efficiency Evaluation of Multi-user Multi-antenna Random Cellular Networks with Minimum Distance Constraints\ldots}
\title{Energy Efficiency of Multi-user Multi-antenna Random Cellular Networks with Minimum Distance Constraints}
\author{Xiaohu~Ge,~\IEEEmembership{Senior~Member,~IEEE,}
        Bangzheng Du,
        Qiang Li,
        and Diomidis S. Michalopoulos~\IEEEmembership{Senior~Member,~IEEE}

\thanks{\scriptsize{Copyright (c) 2015 IEEE. Personal use of this material is permitted. However, permission to use this material for any other purposes must be obtained from the IEEE by sending a request to pubs-permissions@ieee.org.}}
\thanks{\scriptsize{Xiaohu~Ge, Bangzheng~Du, and Qiang~Li (corresponding author) are with the School of Electronic Information and Communications, Huazhong University of Science and Technology, Wuhan 430074, Hubei, P. R. China (email: xhge@mail.hust.edu.cn; bangzhengdu@gmail.com; qli\_patrick@hust.edu.cn).}}
\thanks{\scriptsize{D. S. Michalopoulos is with Nokia Bell Labs, Munich, Germany (email: diomidis.michalopoulos@nokia.com).}}
\thanks{\scriptsize{The authors would like to acknowledge the support from the National Natural Science Foundation of China (NSFC) under the grants 61301128 and 61461136004, NFSC Major International Joint Research Project under the grant 61210002, the Fundamental Research Funds for the Central Universities under the grant 2015XJGH011. This research is partially supported by the EU FP7-PEOPLE-IRSES, project acronym S2EuNet (grant no. 247083), project acronym WiNDOW (grant no. 318992) and project acronym CROWN (grant no. 610524). This research is supported by the National international Scientific and Technological Cooperation Base of Green Communications and Networks (No. 2015B01008) and Hubei International Scientific and Technological Cooperation Base of Green Broadband Wireless Communications.}}
}
\maketitle

\begin{abstract}
Compared with conventional regular hexagonal cellular models, random cellular network models resemble real cellular networks much more closely. However, most studies of random cellular networks are based on the Poisson point process and do not take into account the fact that adjacent base stations (BSs) should be separated with a minimum distance to avoid strong interference among each other. In this paper, based on the hard core point process (HCPP), we propose a multi-user multi-antenna random cellular network model with the aforementioned minimum distance constraint for adjacent BSs. Taking into account the effects of small scale fading and shadowing, interference and capacity models are derived for the multi-user multi-antenna HCPP random cellular networks. Furthermore, a spectrum efficiency model as well as an energy efficiency model is presented, based on which, the maximum achievable energy efficiency of the considered multi-user multi-antenna HCPP random cellular networks is investigated. Simulation results demonstrate that the energy efficiency of conventional Poison point process (PPP) cellular networks is underestimated when the minimum distance between adjacent BSs is ignored.

\end{abstract}

\section{Introduction}
\label{sec1}
\IEEEPARstart{W}{ith} the rapid growth of wireless traffic over the last decade, multiple-input multi-output (MIMO) antenna technology has been widely adopted to satisfy the high traffic requirement in the fourth generation (4G) and future fifth generation cellular networks \cite{Zhang_LTE,Ge16}. On the other hand, the energy consumption in cellular networks has been increasing dramatically because of the increasing number of antennas and the increasing wireless traffic\cite{Ge15}. By 2011, there were more than 4 millions of base stations (BSs) operating in cellular networks, each consuming an average of 25 MWh per year \cite{Hasan_Green}. Therefore, it is important to investigate and improve the energy efficiency of multi-user multi-antenna cellular networks.

 Numerous energy efficiency models for MIMO wireless communication systems have been proposed in the literature \cite{Heliot_An,Heliot_On,Liu_Energy,Belmega_An,Xu_Improving,Xu_Energy,Miao_On,Ngo_Energy,Chen_On}. A closed-form approximation for the energy efficiency-spectrum efficiency trade-off has been derived for the MIMO Rayleigh fading channel in \cite{Heliot_An}. The simulation results in \cite{Heliot_An} indicated that the energy efficiency can be effectively improved through receive diversity in the very low spectrum efficiency regime and that MIMO systems are more energy efficient than single-input single-output (SISO) systems in the high spectrum efficiency regime. Furthermore, the energy efficiency gain of MIMO over SISO systems was analyzed for various power consumption models at transmitters in \cite{Heliot_On}. The MIMO transmission energy efficiency was analyzed for wireless sensor networks considering both the diversity gain and the multiplexing gain in \cite{Liu_Energy}. A precoding matrix was optimized to maximize the energy efficiency of wireless communication systems for single-user MIMO channels in \cite{Belmega_An}. The energy efficiency optimization was investigated by adaptively adjusting the bandwidth, transmission power, and precoding mode in downlink MIMO systems in \cite{Xu_Improving}. Optimizations on the transmit covariance precoding matrix and active transmit antenna selection were proposed to improve the energy efficiency for MIMO broadcast channels in \cite{Xu_Energy}. Based on the distributed singular value decomposition (SVD) of multi-user channels, a power allocation scheme was presented to achieve the optimal energy efficiency in multi-user MIMO systems in \cite{Miao_On}. The trade-off between energy efficiency and spectrum efficiency was quantified for very-large multi-user MIMO systems with small scale fading channels in \cite{Ngo_Energy}. However, most energy efficiency studies of MIMO systems have focused on the link level and single cell scenarios. The case of multi-cell MIMO was also studied in the literature (see, e.g. \cite{Ngo_Energy}, where the conventional regular hexagonal cell model was adopted). Nonetheless, the investigation of the energy efficiency of realistic multi-user multi-antenna networks by following the well-known stochastic geometry approach still remains unexplored.

It is well known that the locations of the transmitters and receivers are very important for the performance of wireless communication systems. In the literature, there are published works which consider a location model in random cellular networks. In \cite{Chen_EMC} and \cite{Chen_AIWAC}, a QoE-driven approach based on a novel mobile cloud computing architecture is proposed to extensively improve the energy efficiency of wireless communication systems. The most popular random process used for wireless network models is the Poisson point process (PPP) \cite{ElSawy_Stochastic}. Pioneering results on random wireless networks were reported in \cite{Andrews_A,Ge14}. The detailed mathematical definition and the properties of the PPP in random wireless networks were discussed in \cite{Chan_Calculating}. In \cite{Yu_Downlink}, a density of success transmissions was derived for the downlink cellular network where the locations of the BSs were governed by a homogeneous PPP. Based on the stochastic geometry and the PPP theory, a simple single integral model for the average rate of random cellular networks was derived in \cite{Guidotti_Simplified}, which is useful for performance analysis. In \cite{DiRenzo_Average}, a comprehensive mathematical framework was proposed for the analysis of the average rate of multi-tier cellular networks whose BSs are assumed to follow a PPP distribution. A signal-to-interference-plus-noise ratio (SINR) model was derived for multi-tier cellular networks where the locations of the BSs followed a PPP \cite{Mukherjee_Distribution}. For Poisson distributed multi-antenna BSs, an approximation for the area-averaged spectral efficiency of a representative link was derived in \cite{Govindasamy_Asymptotic}. Assuming that each tier of BSs was modeled by an independent homogeneous PPP, a tractable downlink model for multi-antenna heterogeneous cellular networks was proposed in \cite{Dhillon_Downlink}. Adopting the PPP for the distribution of the BSs, success probability and energy efficiency models for homogeneous single-tier macrocell and heterogeneous multi-tier small cell networks were derived under different sleeping policies in \cite{Soh_Energy}.

In addition to the PPP, other random point processes have also been used for modeling and performance analysis of wireless networks in \cite{Srinivasa_Modeling,Ganti_Interference,ElSawy_Characterizing,Guo_Spatial}. For a wireless network with a finite and fixed number of nodes, a closed-form analytical expression for the moment generation function of the interference was derived based on the binomial point process (BPP) in \cite{Srinivasa_Modeling}. For the case where node locations of clustered wireless ad hoc networks are assumed to form a Poisson cluster process (PCP), in \cite{Ganti_Interference} the distribution properties of interference were derived for analyzing the outage probability. Furthermore,
considering a minimal distance constraint between nodes in random carrier-sense multiple access (CSMA) wireless networks, a modified hard core point process (HCPP) was used for modeling the spatial distribution of the simultaneously active users in \cite{ElSawy_Characterizing}. Different spatial stochastic models including the PPP, the HCPP, the Strauss process (SP) and the perturbed triangular lattice were compared for modeling the spatial distribution of BSs in cellular networks and it was proven that the HCPP is more realistic than the PPP for modeling the spatial structure of BSs \cite{Guo_Spatial}. However, a detailed investigation of the performance of multi-user multi-antenna cellular networks under the HCPP is not available in the literature.
Therefore, motivated by the above gaps in the literature, in this paper, we derive the average energy efficiency of multi-user multi-antenna cellular networks with HCPP distributed BSs. The contributions of this paper are summarized as follows.

\begin{enumerate}
\item Considering the minimum distance constraint of adjacent BSs, a stochastic spatial distribution for the BSs is proposed for multi-user multi-antenna cellular networks, based upon the HCPP. The HCPP was traditionally used for wireless local area networks (WLANs) without interference. In this paper, the HCPP cellular scenario is proposed for describing the minimum distance constraint of adjacent BSs in random cellular networks compared with traditional PPP random cellular networks.
\item We propose a model for calculating the average interference in multi-user multi-antenna random cellular networks with HCPP distributed interfering transmitters. Numerical results indicate that the average interference is underestimated when the minimum distance in adjacent BSs is ignored.
\item For the proposed interference model and the zero-forcing precoding at BSs, spectrum efficiency and energy efficiency models are derived for multi-user multi-antenna HCPP cellular networks.
\item Simulation results illustrate how the maximal energy efficiency of multi-user multi-antenna HCPP cellular networks depends on the number of antennas, the wireless traffic distribution, the propagation of the wireless channel, and the minimum distance between adjacent BSs. Based on simulation comparisons, it is implied that the energy efficiency of conventional PPP cellular networks is underestimated when the minimum distance in adjacent BSs is ignored.
\end{enumerate}

The remainder of this paper is organized as follows. Section II introduces the system model. In Section III, an average interference model is derived for multi-antenna HCPP cellular networks. Assuming zero-forcing precoding at the BSs, a model for evaluating the spectrum efficiency of multi-user multi-antenna HCPP cellular networks is proposed in Section IV. Furthermore, a model for the energy efficiency of multi-user multi-antenna HCPP cellular networks is presented in Section V.  Simulation and analytical results are presented in Section VI. Finally, conclusions are drawn in Section VII.

\section{System model}
\label{sec2}
Compared with the regular hexagonal cell structure, random cellular networks are more coincident with real deployments of cellular networks. The PPP theory has been widely used for modelling of random cellular networks \cite{Andrews_A}. However, there is no constraint for the distance between two points in PPPs. As a consequence, there exist scenarios where two adjacent BSs are infinitesimally close to each other  for PPP random cellular networks. In this case, the interference from adjacent BSs will approach to infinity when the interfering BSs are infinitesimally close to the desired BS in a PPP random cellular network. In realistic BSs deployments in cellular networks, two arbitrary BSs cannot be infinitesimally closed to each other. In general, telecommunication providers always ask that the location of two adjacent BSs must keep a protect distance or a minimum distance to avoid obvious interference. Hence, there exists a conflict for the minimum distance constraint between two arbitrary BSs in realistic BSs deployment and PPP random cellular networks. This result will conduce to the deviation for interference and energy efficiency analysis for random cellular networks, which are illustrated in Figs. 2 and 11. To solve this drawback of PPP random cellular network, we propose to model random cellular network based on the HCPP theory. In the following, we introduce the channel model, the traffic model and the HCPP, which has been shown to be more realistic in modeling the deployments of cellular networks than the PPP.
\subsection{Overview of HCPP}

In the literature, the PPP is widely used for modelling random cellular networks, mainly because it leads to a mathematically tractable analysis \cite{Yu_Downlink,Guidotti_Simplified,DiRenzo_Average,Mukherjee_Distribution,Govindasamy_Asymptotic,Dhillon_Downlink,Soh_Energy,Win_A}. However, some aspects associated with the PPP analysis may render it inadequate for modelling certain realistic cellular deployments. For example, in downlink interference models of PPP random cellular networks, the locations of the interfering BSs can approach that of the desired BS arbitrarily close. As a result, the mean of the aggregated interference approaches infinity \cite{Win_A}. This result not only increases the model complexity but also deviates from reality. To avoid this extreme result, we base our random cellular network model on the HCPP theory \cite{Haenggi_Mean,Matern_Spatial,Stoyan_Stochastic}. We note that this theory has been used for the modeling of carrier sense multiple access (CSMA) networks with a specified minimum distance between adjacent wireless nodes \cite{ElSawy_Modeling}. HCPP generates patterns produced by points that have a minimum distance $\delta $ from each other. The Matern hard-core process of Type II, which represents a special case of HCPP, is essentially a stationary PPP ${\prod _{{\text{PPP}}}}$ , i.e., the Poisson point process of intensity ${\lambda _P}$, to which a dependent thinning is applied \cite{Stoyan_Stochastic}. The thinned process ${\prod _{{\text{HCPP}}}}$, i.e., the Mat¨¦rn hard-core process is expressed as \cite{Stoyan_Stochastic}
\[\begin{gathered}
  {\Pi _{{\text{HCPP}}}} \hfill \\
   = \{ x \in {\Pi _{{\text{PPP}}}}:\Phi (x) < \Phi ({x^ * }){\text{ for all }}{x^ * }{\text{ in }}{\Pi _{{\text{PPP}}}} \cap d(x,\delta )\}  \hfill \\
\end{gathered}.\tag{1} \]
The points of ${\prod _{{\text{PPP}}}}$ are marked with random numbers uniformly distributed in $[0,1]$ independently. The dependent thinning retains the point $x$ of ${\prod _{{\text{PPP}}}}$ with mark $\Phi (x)$ if the disk $d(x,\delta )$ contains no points of ${\prod _{{\text{PPP}}}}$ with marks smaller than $\Phi (x)$, where $d(x,\delta )$ is a disk area with central point $x$ and the radius ${\delta}$.

We assume that both BSs and user equipments (UEs) are randomly located in the infinite plane $\mathbb{R}^2$. Moreover, the UEs¡¯ motions are isotropic and relatively slow, such that during an observation period, e.g., a time slot, the relative positions of BSs and UEs are stationary. The distribution of the BSs is assumed to be governed by a thinned process ${\prod _{{\text{HCPP}}}}$ applied to a stationary PPP ${\prod _{{\text{PPP}}}}$ of intensity ${\lambda _P}$. The locations of the BSs are denoted by ${\prod _{BS}} = \{ {x_{BS_i}}:i = 1,2,3 \cdots \} $, where ${x_{BS_i}}$ are the two-dimensional Cartesian coordinates that denote the location of BS $BS_i$. The distribution of UEs is assumed to be governed by a PPP with intensity ${\lambda _M}$.

\subsection{Channel Model}
We assume that the BS and UE are equipped with ${N_T}$ and ${N_R}$ antennas, respectively. We also assume that each UE connects to the closest BS, which corresponds to the smallest path loss during wireless transmission. In this paper, our studies are focused on the downlink of random cellular networks. The large scale fading coefficients of the UEs in a cell are assumed to be identical to each other. The channel matrix between UEs and BS is modelled as
\[{\bf{H}} = \sqrt {\frac{\beta }{{{\Re ^\alpha }}}w} {\bf{h}}, \ \tag{2}\]
where ${\mathbf{H}}$ is the channel matrix, $\beta $ is a constant depending on the antenna gain, $\Re$ is the distance between the transmitter and receiver, and $\alpha$ is the path loss coefficient. Furthermore, $w$ models the log-normal shadowing effect in wireless channels and is given by $w = {10^{s/10}}$, where $s$ is a Gaussian distributed random variable with zero mean and variance $\sigma _s^2$ , $s \sim N(0,\sigma _s^2)$. Additionally, ${\bf{h}}$ is the small scale fading channel matrix, whose elements are modelled as independent and identically distributed \emph{(i.i.d.)} Gaussian random variables with zero mean and unit variance.

\subsection{Traffic Model}
In early studies \cite{Frost94}, the Poisson distribution was adopted for traffic modelling of cellular networks. Based on empirical measurement results \cite{Lilith_Using}, the traffic load of cellular networks has been demonstrated to have the self-similar characteristic which means the variance of similar network traffic approaches to infinity. To model the cellular network traffic with self-similar characteristic, several mathematical distributions with the infinite variance have been proposed to fit the self-similar network traffic \cite{Silva_Performance,Norros95,Karasaridis01}. Considering the analytical expression and intuitionistic engineering implication of function parameters, e.g., the traffic rate $\rho $, the Pareto distribution has been widely used for similar cellular network traffic modelling \cite{Xiang_Energy}. Without loss of generality, the Pareto distribution has been adopted for the cellular network traffic in this study. Moreover, the traffic rates of all UEs are assumed to be \emph{i.i.d.} The probability density function (PDF) of traffic rate in cellular networks is given by
\[{f_\rho }(\chi ) = \frac{{\theta \rho _{\min }^\theta }}{{{\chi ^{\theta  + 1}}}},\chi  \ge {\rho _{\min }} > 0,\ \tag{3}\]
where $\theta  \in (1,2]$ is the heaviness index which reflects the heaviness of the distribution tail, and ${\rho _{\min }}$ denotes the minimum traffic rate which is configured to guarantee the user requirement in data transmission rate. We note that the heaviness index, $\theta$, affects the distribution tail of the traffic rate, so that when $\theta$ approaches 1 the tail becomes the dominant part of the distribution. The average traffic rate at UEs is obtained as
\[\mathbb{E}(\rho ) = \frac{{\theta {\rho _{\min }}}}{{\theta  - 1}},\ \tag{4}\]
where $\mathbb{E}( \cdot )$ denotes the expectation operation.

\section{Interference Model of HCPP Cellular Networks}
\label{sec3}
To evaluate wireless propagation environments, an average interference model has been proposed for HCPP cellular networks in this section. Moreover, the impacts of the distance between the UE and desired BS, the minimum BSs distance and path loss coefficient on the average interference of HCPP cellular networks are analyzed by numerical simulations.

\subsection{Interference Model}

\begin{figure*}[!t]
\vspace{0.1in}
\centerline{\includegraphics[width=15cm,draft=false]{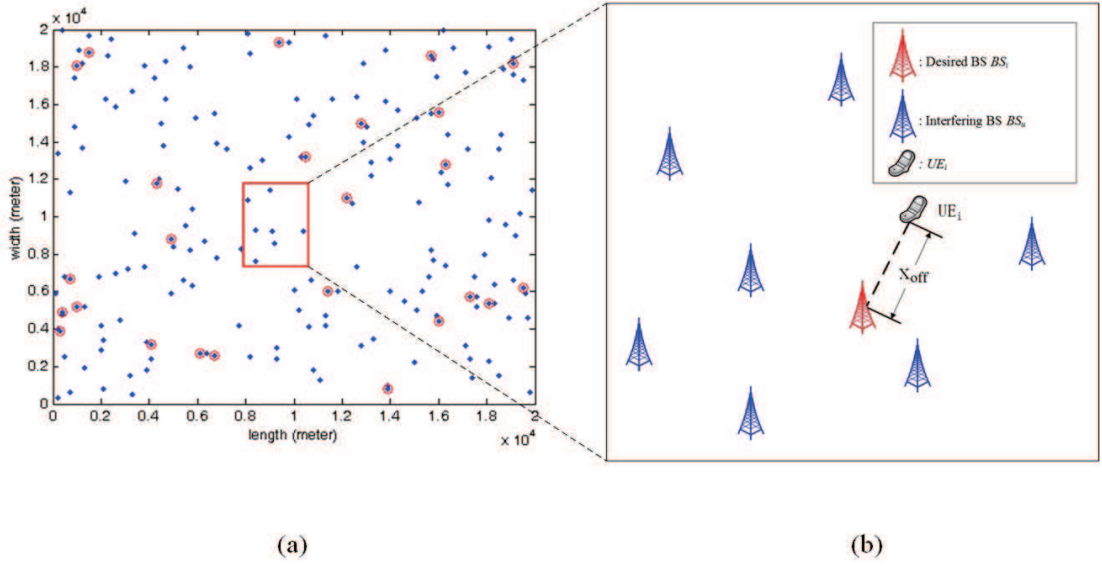}}
\caption{\small Illustration of HCPP BSs distribution.}
\end{figure*}

In Fig. 1(a), based on the HCPP model, a simulation-based illustration of the distribution of the BSs for the considered system is shown. The blue nodes represent BSs and are distributed according to a PPP with intensity $\lambda _P = 1/(\pi *800^2)$. The minimum distance is set to $\delta  = 500$ meter. The nodes whose marked values do not satisfy the condition in (1) are marked by red circles. As mentioned in the HCPP model in section II, these red circle points are discarded from the analysis. As a consequence, the BSs that are included in the HCPP model are the nodes marked with blue, whose distance from adjacent nodes is larger than or equal to $\delta$.

In a hard core point distribution, a point with mark $t$, $t \in \left[ {0,1} \right]$, is retained only when there are no other points with a smaller mark at a distance less than the hard core distance $\delta $. In a Poisson point distribution, given that the mean of point number is ${\lambda _P}\pi {\delta ^2}$ in a circle with radius $\delta $, the probability of  points is expressed as
\[\begin{gathered}
  {\text{Pr}}\left\{ {{\text{poisson point number in }}\pi {\delta ^2}{\text{ area is }}{\kern 1pt} \kappa } \right\} \hfill \\
   = \frac{{{{({\lambda _P}\pi {\delta ^2})}^\kappa }{e^{ - {\lambda _P}\pi {\delta ^2}}}}}{{\kappa !}} \hfill \\
\end{gathered}.\tag{5} \]

The probability that the point with mark $t$ is retained in a Poisson point distribution is derived as
\[\begin{gathered}
  {\text{Pr}}\left\{ {\text{A point with mark }}t{\text{ is retained}}\right\} \hfill \\
   = \sum\limits_{\kappa  = 0}^\infty  {\frac{{{{({\lambda _P}\pi {\delta ^2})}^\kappa }{e^{ - {\lambda _P}\pi {\delta ^2}}}}}{{\kappa !}}} {(1 - t)^\kappa } \hfill \\
   = \sum\limits_{\kappa  = 0}^\infty  {\frac{{{{[{\lambda _P}\pi {\delta ^2}(1 - t)]}^\kappa }{e^{ - {\lambda _P}\pi {\delta ^2}}}}}{{\kappa !}}}  = {e^{ - {\lambda _P}\pi {\delta ^2}t}} \hfill \\
\end{gathered}. \ \tag{6}\]

As a result, the retaining probability for a typical point can be calculated by integrating ${e^{ - {\lambda _P}\pi {\delta ^2}t}}$ in the interval $t \in \left[ {0,1} \right]$. At the location ${x}$, the probability that there is a point in the infinitesimal small region $dx$ is ${\zeta ^{(1)}}dx$. The first moment of HCPP is expressed by
\[{\zeta ^{(1)}} = {\lambda _P}\int_0^1 {{e^{ - {\lambda _P}\pi {\delta ^2}t}}dt}  = \frac{{1 - {e^{ - {\lambda _P}\pi {\delta ^2}}}}}{{\pi {\delta ^2}}}.\ \tag{7}\]

When two points with mark ${t_1}$ and ${t_2}$ are located at two differential regions $d{{x}_1}$ and $d{{x}_2}$ in a PPP distribution, the probability that two points are retained depends only on the distance $r$ between two points. Moreover, if two circles with the same radius $\delta$ are separated by $r$, the area of the union of two circles ${V_\delta }(r)$ is derived as \cite{Haenggi_Mean}
\[{V_\delta }(r) = \left\{ {\begin{array}{*{20}{c}}
  {2\pi {\delta ^2} - 2{\delta ^2}\arccos (\frac{r}{{2\delta }}) + r\sqrt {{\delta ^2} - \frac{{{r^2}}}{4}} ,2\delta  > r > 0} \\
  \qquad\qquad\qquad\qquad\qquad\quad{2\pi {\delta ^2},r \geqslant 2\delta }
\end{array}} \right..\ \tag{8}\]

Using (6) and (7), the probability that two points with stamp marks ${t_1}$ and ${t_2}$ are retained can be derived by considering the case of $r > \delta$ and $r \leqslant \delta$ independently,
\[\begin{gathered}
  \varphi (r) = \int_0^1 {{e^{ - {\lambda _P}{t_1}\pi {\delta ^2}}}} \int\limits_0^{{t_1}} {{e^{ - {\lambda _P}{t_2}[{V_\delta }(r) - \pi {\delta ^2}]}}d{t_2}d{t_1}}  \hfill \\
  {\kern 1pt} {\kern 1pt} {\kern 1pt} {\kern 1pt} {\kern 1pt} {\kern 1pt} {\kern 1pt} {\kern 1pt} {\kern 1pt} {\kern 1pt} {\kern 1pt} {\kern 1pt} {\kern 1pt} {\kern 1pt} {\kern 1pt} {\kern 1pt} {\kern 1pt} {\kern 1pt} {\kern 1pt} {\kern 1pt} {\kern 1pt} {\kern 1pt} {\kern 1pt} {\kern 1pt} {\kern 1pt}  + \int_0^1 {{e^{ - {\lambda _P}{t_2}\pi {\delta ^2}}}} \int\limits_0^{{t_2}} {{e^{ - {\lambda _P}{t_1}[{V_\delta }(r) - \pi {\delta ^2}]}}d{t_1}d{t_2}}  \hfill \\
  {\text{       }} = \left\{ {\begin{array}{*{20}{c}}
  {\frac{{2{V_\delta }(r)(1 - {e^{ - {\lambda _P}\pi {\delta ^2}}}) - 2\pi {\delta ^2}(1 - {e^{ - {\lambda _P}{V_\delta }(r)}})}}{{\lambda _P^2\pi {\delta ^2}{V_\delta }(r)[{V_\delta }(r) - \pi {\delta ^2}]}},r > \delta } \\
  {0,r \leqslant \delta }
\end{array}} \right. \hfill \\
\end{gathered}.\tag{9} \]
Furthermore, the second moment of HCPP is expressed by
\[{\zeta ^{(2)}}(r) = \lambda _p^2\varphi (r).\ \tag{10}\]
As a consequence, the probability that the distance between two points which are located in the infinitesimally small regions, $d{{x}_1}$ and $d{{x}_2}$, respectively, equals $r$, is expressed as ${\zeta ^{(2)}}(r)d{{x}_1}d{{x}_2}$.

Let us denote the desired BS and UE by $BS_i$ and $UE_i$, respectively. Based on the system model in Fig. 1(b), the location vector of desired BS $BS_i$ is denoted by ${{x}_{BS_i}}$, the distance vector between $UE_i$ and $BS_i$ is denoted by ${{x}_{off}}$. Moreover, ${{x}_{BS_u}}$ is the two-dimensional Cartesian coordinate of interfering BS, denoted by $BS_u$. Considering the impact of the distance between the desired BS and the received UE \cite{Andrews07}, the aggregated interference at $UE_i$ is expressed as
\[{I_i}({{x}_{off}}) = \sum\limits_{u \ne i,BS_u \in {\Pi _{BS}}} {g(||{{x}_{BS_u}} - {{x}_{BS_i}} - {{x}_{off}}||,{\psi _{_{iu}}})}, \ \tag{11}\]
where $g$ is defined as the interference function between ${BS}_u$ and $UE_i$, $|| \cdot ||$ is the modular operation, i.e., the Euclid distance operation. Furthermore, ${\psi _{iu}}$ is the fading factor over wireless channels, given as ${\psi _{iu}} = \{ {w_{iu}},{{\mathbf{h}}_{iu}},{P_u}\} $, which includes the shadowing effect ${w_{iu}}$, the small scale fading matrix ${{\mathbf{h}}_{iu}}$ and the transmission power ${P_u}$ at $B{S_u}$. Based on the channel model introduced in Section II, the aggregated interference at $UE_i$ can be extened as
\[{I_i}({{x}_{off}}) = \sum\limits_{u \ne i,BS_u \in {\Pi _{BS}}} {\frac{{\beta {w_{iu}}|{z_{iu}}{|^2}{P_u}}}{{||{{x}_{BS_u}} - {{x}_{BS_i}} - {{x}_{off}}|{|^\alpha }}}}, \ \tag{12}\]
where ${z_{iu}}$ is the small scale fading between the received UE $U{E_i}$ and the interfering BS $B{S_u}$ with single antenna and is governed by a complex Gaussian distribution with mean value that equals 1.

Considering every active UE is associated with a BS and all UEs are traversed in the plane $\mathbb{R}^2$, the total interference in the plane $\mathbb{R}^2$ is given by
\[\begin{gathered}
  \sum\limits_{B{S_i} \in {\Pi _{BS}}} {{I_i}({x_{off}})}  \hfill \\
   = \sum\limits_{B{S_i} \in {\Pi _{BS}}} {\sum\limits_{u \ne i,B{S_u} \in {\Pi _{BS}}} {g(||{x_{B{S_u}}} - {x_{B{S_i}}} - {x_{off}}||,{\psi _{_{iu}}})} }  \hfill \\
\end{gathered}.\tag{13} \]

Based on the second moment of the HCPP in (10) and the corresponding properties \cite{Byungjin_Bounding}, the expectation of the total interference in the plane $\mathbb{R}^2$ is derived as
\[\begin{gathered}
  \mathbb{E}[\sum\limits_{B{S_i} \in {\Pi _{BS}}} {{I_i}({x_{off}})} ] \hfill \\
   = \int\limits_{{\mathbb{R}^2}} {\int\limits_{{\mathbb{R}^2}} {\left\{ \begin{gathered}
  {E_{{\psi _{iu}}}}[g(||{x_1} - {x_2} - {x_{off}}||,{\psi _{iu}})] \hfill \\
   \times {\zeta ^{(2)}}(||{x_1} - {x_2} - {x_{off}}||) \hfill \\
\end{gathered}  \right\}d{x_1}d{x_2}} }  \hfill \\
\end{gathered}.\tag{14} \]

Based on the first moment of HCPP in (7), the average BS number in the plane $\mathbb{R}^2$ is expressed as $\int\limits_{\mathbb{R}^2} {{\zeta ^{(1)}}dx} $. Without loss of generality, the aggregated interference $\sum\limits_{B{S_i} \in {\prod _{BS}}} {{I_i}\left( {{x_{off}}} \right)} $ can be calculated by the distance among the BSs $\left\| {{x_{B{S_u}}} - {x_{B{S_i}}}} \right\|$ and the distance between the desired BS and the received UE $\left\| {{x_{off}}} \right\|$. Let $x = {x_{B{S_u}}} - {x_{B{S_i}}}$ be the distance vector among the BSs in HCPP cellular networks, the average interference of HCPP cellular networks is derived as
\[\begin{gathered}
  {I_{i\_avg}} \hfill \\
   = \frac{{\int\limits_{{\mathbb{R}^2}} {\int\limits_{{\mathbb{R}^2}} {\left\{ \begin{gathered}
  {\mathbb{E}_{{\psi _{iu}}}}[g(||{x_1} - {x_2} - {x_{off}}||),{\psi _{iu}}] \hfill \\
   \times {\varsigma ^{(2)}}(||{x_1} - {x_2} - {x_{off}}||) \hfill \\
\end{gathered}  \right\}d{x_1}d{x_2}} } }}{{\int\limits_{{\mathbb{R}^2}} {{\varsigma ^{(1)}}dx} }} \hfill \\
   = \frac{{\int\limits_{{\mathbb{R}^2}} {d{x_1}\int\limits_{{\mathbb{R}^2}} {\left\{ \begin{gathered}
  {\mathbb{E}_{{\psi _{iu}}}}[g(|| - {x_2} - {x_{off}}||),{\psi _{iu}}] \hfill \\
   \times {\varsigma ^{(2)}}(|| - {x_2} - {x_{off}}||) \hfill \\
\end{gathered}  \right\}d{x_2}} } }}{{{\varsigma ^{(1)}}\int\limits_{{\mathbb{R}^2}} {dx} }} \hfill \\
   = \frac{1}{{{\varsigma ^{(1)}}}}\int\limits_{{\mathbb{R}^2}} {{\mathbb{E}_{{\psi _{iu}}}}[g(||x + {x_{off}}||),{\psi _{iu}}]{\varsigma ^{(2)}}(||x + {x_{off}}||)dx}  \hfill \\
\end{gathered}.\tag{15} \]


Substituting (11) and (12) into (15), the average interference of HCPP cellular networks is derived as
\[\begin{gathered}
  {I_{i\_avg}}({x_{off}}) = \frac{{\beta \mathbb{E}({w_{iu}})\mathbb{E}(|{z_{iu}}{|^2})\mathbb{E}({P_u})}}{{{\zeta ^{(1)}}}} \hfill \\
  {\kern 1pt} {\kern 1pt} {\kern 1pt} {\kern 1pt} {\kern 1pt} {\kern 1pt} {\kern 1pt} {\kern 1pt} {\kern 1pt} {\kern 1pt} {\kern 1pt} {\kern 1pt} {\kern 1pt} {\kern 1pt} {\kern 1pt} {\kern 1pt} {\kern 1pt} {\kern 1pt} {\kern 1pt} {\kern 1pt} {\kern 1pt} {\kern 1pt} {\kern 1pt} {\kern 1pt} {\kern 1pt} {\kern 1pt} {\kern 1pt} {\kern 1pt} {\kern 1pt} {\kern 1pt} {\kern 1pt} {\kern 1pt} {\kern 1pt} {\kern 1pt} {\kern 1pt} {\kern 1pt} {\kern 1pt} {\kern 1pt} {\kern 1pt} {\kern 1pt} {\kern 1pt} {\kern 1pt} {\kern 1pt} {\kern 1pt} {\kern 1pt} {\kern 1pt} {\kern 1pt} {\kern 1pt} {\kern 1pt} {\kern 1pt} {\kern 1pt} {\kern 1pt} {\kern 1pt} {\kern 1pt} {\kern 1pt} {\kern 1pt} {\kern 1pt} {\kern 1pt} {\kern 1pt} {\kern 1pt}  \times \int\limits_{{\mathbb{R}^2}} {\frac{1}{{||x + {x_{off}}|{|^\alpha }}}{\zeta ^{(2)}}(||x||)dx}  \hfill \\
\end{gathered},\tag{16} \]
where the distance ${{x}_{off}}$ between the transmitter and the receiver is considered to evaluate the average interference of HCPP cellular networks.

\subsection{Performance Analysis of Interference Model}
Based on (16), the impacts of the distance between the user and the desired BS, the path loss coefficient and the minimum distance on the average interference of HCPP cellular networks are numerically analyzed in detail. In the following analysis, the default parameters used for interference model are configured as follows: ${\sigma _s} = 6$, which usually ranges from 4 to 9 in practice \cite{Simon_Digital}; $\alpha  = 3.8$ and $\beta  =  - 31.54$ dB, which correspond to an urban area with a rich scattering environment \cite{Xiang_Energy}; the average transmission power of interfering BS is set as $\mathbb{E}({P_u}) = 2$ Watt (W) when the interference link bandwidth is configured as 10 KHz \cite{Cui_Energy,Arnold_Power}. Moreover, analysis results are confirmed by Monte-Carlo (MC) simulations in HCPP cellular networks.

\begin{figure}
\centering
\subfigure[\small Average interference in HCPP model]
{
\label{fig:subfig:a}
\includegraphics[width=9.5cm,draft=false]{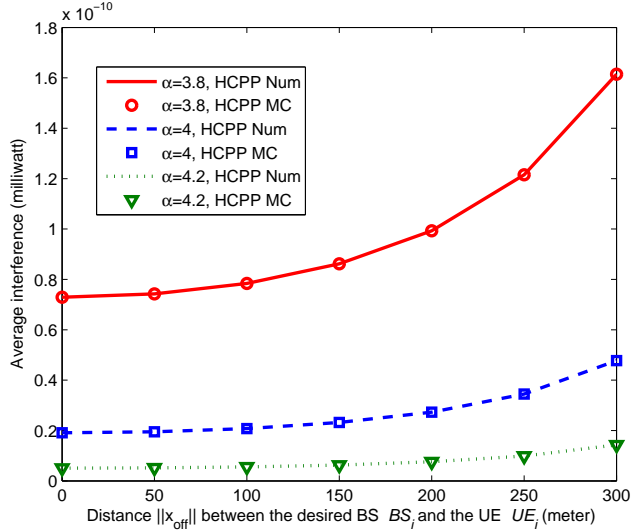}
}
\hspace{0.5cm}
\subfigure[\small Average interference in PPP model]
{
\label{fig:subfig:b}
\includegraphics[width=9.5cm,draft=false]{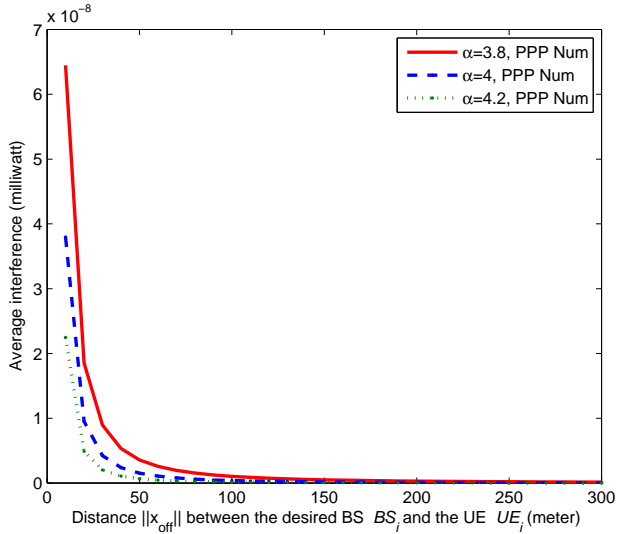}
}
\caption{\small Impact of the distance $\left\| {{{x}_{off}}} \right\|$ between the desired BS, $B{S_i}$, and the UE of interest, $U{E_i}$, on the average interference of HCPP cellular networks.}
\label{fig:subfig}
\end{figure}

Fig. 2 illustrates the impact of the distance $\left\| {{{x}_{off}}} \right\|$ between the desired BS $BS_i$ and the UE $UE_i$ on the average interference of HCPP and PPP cellular networks. When the path loss coefficient $\alpha $ is fixed, the average interference of HCPP cellular networks increases with increasing the distance $\left\| {{{x}_{off}}} \right\|$ and the average interference of PPP cellular networks decreases with increasing the distance $\left\| {{{x}_{off}}} \right\|$. When the distance $\left\| {{{x}_{off}}} \right\|$ between the desired BS $BS_i$ and the UE $UE_i$ is fixed, both HCPP and PPP cellular networks indicate that the average interference increases with decreasing the path loss coefficient $\alpha $ .

\begin{figure}
\vspace{0.1in}
\centerline{\includegraphics[width=10cm,draft=false]{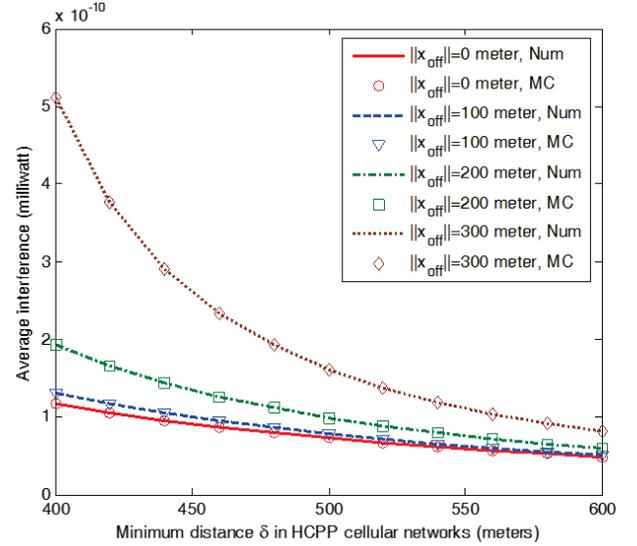}}
\caption{\small Impact of the minimum distance $\delta $ on the average interference of HCPP cellular networks.}
\end{figure}

Fig. 3 shows the impact of the minimum distance on the average interference of HCPP cellular networks, in which ¡°Num¡± labels numerical results and ¡°MC¡± represents MC simulation results. When the distance $\left\| {{{x}_{off}}} \right\|$ between the desired BS $BS_i$ and the UE $UE_i$ is fixed, the average interference decreases with increasing the minimum distance $\delta $ in HCPP cellular networks. When the minimum distance  is fixed, the average interference increases with increasing the distance $\left\| {{{x}_{off}}} \right\|$ in HCPP cellular networks. Based on simulation results in Fig. 3, the average interference is underestimated when the distance $\left\| {{{x}_{off}}} \right\|$ is ignored in HCPP cellular networks.

\begin{figure}
\vspace{0.1in}
\centerline{\includegraphics[width=10cm,draft=false]{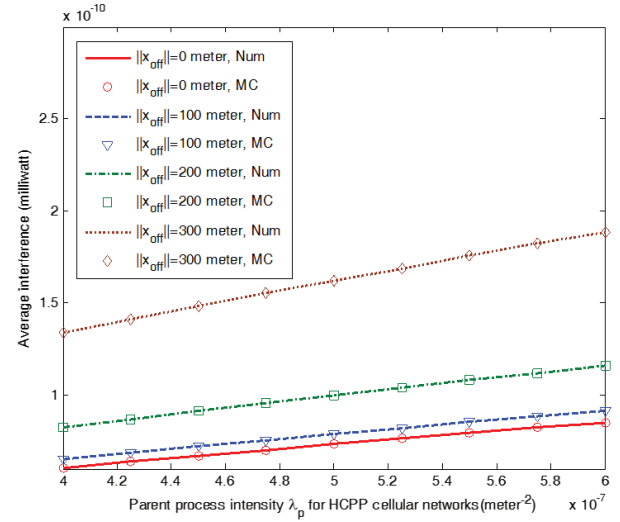}}
\caption{\small Impact of the parent process intensity ${\lambda _P}$ on the average interference of HCPP cellular networks.}
\end{figure}

Fig. 4 analyzes the impact of the parent process intensity ${\lambda _P}$ on the average interference of multi-antenna HCPP cellular network. When the distance $\left\| {{{x}_{off}}} \right\|$ between the desired BS $B{S_i}$ and the UE $U{E_i}$ is fixed, the average interference increases with increasing the parent process intensity ${\lambda _P}$ in multi-antenna HCPP cellular networks. When the parent process intensity ${\lambda _P}$ is fixed, the average interference increases with increasing the distance $\left\| {{{x}_{off}}} \right\|$ between the desired BS $B{S_i}$ and the UE $U{E_i}$ in multi-antenna HCPP cellular networks. Simulation results in Fig. 4 indicate that the average interference is underestimated when the distance $\left\| {{{x}_{off}}} \right\|$ is ignored in HCPP cellular networks.

\section{Spectrum and Energy Efficiency of Multi-user Multi-antenna HCPP Cellular Networks}
\label{sec4}

\subsection{Spectrum Efficiency of Multi-user Multi-antenna HCPP Cellular Networks}

\begin{figure}
\vspace{0.1in}
\centerline{\includegraphics[width=8.5cm,draft=false]{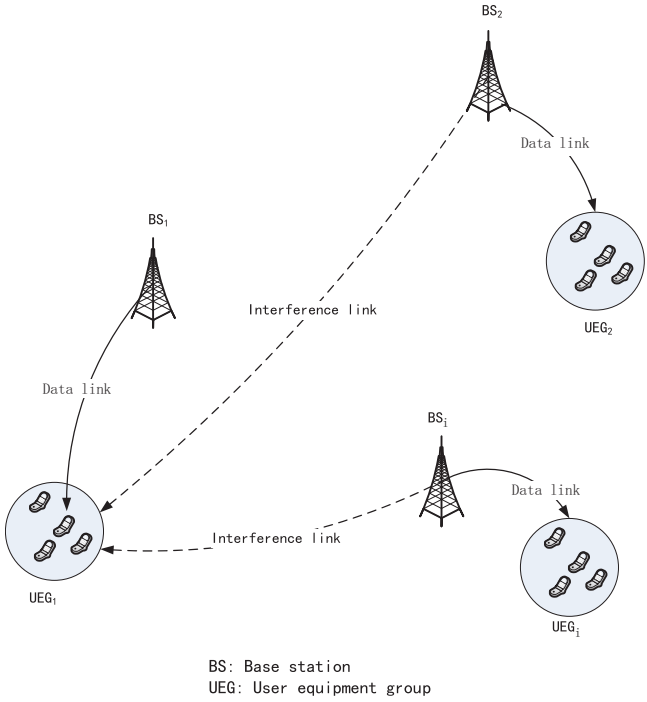}}
\caption{\small Downlink model of multi-user multi-antenna HCPP cellular networks.}
\end{figure}

In this paper, the zero-forcing precoding method is adopted for multi-user multi-antenna downlink systems. Based on the zero-forcing precoding method, the BS integrated with ${N_T}$ antennas can simultaneously transmit signals to $S$ active UEs in a cell, as illustrated in Fig. 5. All  $S$ UEs in a cell are grouped as a user equipment group (UEG). When each UE is equipped with a single antenna, the antenna number of UEG is $S$. We assume that the antenna number of BS is larger than or equal to the antenna number of UEG, i.e., ${N_T} \geqslant S$. The radius of cellular coverage in current cellular networks is about 200-400 meters in the urban regions. When the protect distance, i.e., the minimum distance is configured in HCPP cellular networks, all active USs in a cell are located at a confined circular ring where has an approximated same distance to the desired BS. Considering that the large scale fading depends on the distances between all active UEs in a cell and the desired BS, the large scale fading in a cell can be regarded to be almost identical in HCPP cellular networks. This scenario where UEs in a cell are susceptible to the same large scale fading, was also adopted in \cite{Chen_Performance} and identified as the homogeneous scenario. The location vector of UEG $UE{G_i}$ associated with the BS $B{S_i}$ is denoted as ${{x}_{B{S_i}}}{ + }{{x}_{off}}$.

In Fig. 5, the signal vector ${{\mathbf{y}}_i}$ received by $UE{G_i}$ can be expressed as
\[{{\mathbf{y}}_i} = {{\mathbf{H}}_{ii}}{{\mathbf{x}}_i} + \sum\limits_{u \ne i,B{S_u} \in {\Pi _{BS}}} {{{\mathbf{H}}_{iu}}{{\mathbf{x}}_u}}  + {\mathbf{n}},\ \tag{17a}\]
with
\[{{\mathbf{x}}_i} = {{\mathbf{F}}_i}{{\mathbf{s}}_i},\ \tag{17b}\]
\[{{\mathbf{x}}_u} = {{\mathbf{F}}_u}{{\mathbf{s}}_u},\ \tag{17c}\]
where ${{\mathbf{H}}_{ii}}$ is the channel matrix between $B{S_i}$ and $UE{G_i}$,  ${{\mathbf{x}}_i}$ is the signal vector from $B{S_i}$, ${{\mathbf{H}}_{iu}}$ is the channel matrix between $B{S_u}$ and $UE{G_i}$, ${{\mathbf{x}}_u}$ is the signal vector from $B{S_u}$, and ${\mathbf{n}}$ is the noise vector. The noise power at each antenna is denoted by $\sigma _n^2$, the covariance matrix of noise vectors at $UE{G_i}$ is denoted by $\mathbb{E}({\mathbf{n}}{{\mathbf{n}}^ + }) = \sigma _n^2{{\mathbf{I}}_{S \times S}}$, where ${{\mathbf{I}}_{S \times S}}$ is the $S \times S$ identity matrix. Considering that zero-forcing precoding is adopted for all BSs, ${{\mathbf{F}}_i}$ and ${{\mathbf{F}}_u}$ are the ${N_T} \times S$ precoding matrixes used at $B{S_i}$ and $B{S_u}$, respectively. ${{\mathbf{s}}_i}$ and ${{\mathbf{s}}_u}$ are the $S \times 1$ signal vectors at $B{S_i}$ and $B{S_u}$, respectively.

For zero-forcing precoding, precoding matrix ${{\mathbf{F}}_i}$ can be expressed as
\[{{\mathbf{F}}_i} = {\mathbf{H}}_{ii}^ + {({{\mathbf{H}}_{ii}}{\mathbf{H}}_{ii}^ + )^{ - 1}},\ \tag{18}\]
where ${( \cdot )^ + }$ and ${( \cdot )^{ - 1}}$ denote the conjugate transpose operation and inverse operation, respectively. Furthermore, the signal vector ${{\mathbf{y}}_i}$ received by $UE{G_i}$ can be rewritten as
\[{{\mathbf{y}}_i} = {{\mathbf{s}}_i} + \sum\limits_{u \ne i,B{S_u} \in {\Pi _{BS}}} {{{\mathbf{H}}_{iu}}{{\mathbf{F}}_u}} {{\mathbf{s}}_u} + {\mathbf{n}}.\ \tag{19}\]

Based on the zero-forcing precoding method  \cite{Wang_Scheduling}, the transmission power of $B{S_i}$ is expressed as

\[\begin{gathered}
  {P_i}        = \mathbb{E}({\mathbf{x}}_i^ + {{\mathbf{x}}_i}) \hfill \\
  \;\;\;\; = \mathbb{E}({\mathbf{s}}_i^ + {\mathbf{F}}_i^ + {{\mathbf{F}}_i}{{\mathbf{s}}_i}) \hfill \\
  \;\;\;\; = \mathbb{E}\{ {\mathbf{s}}_i^ + {[{({{\mathbf{H}}_{ii}}{\mathbf{H}}_{ii}^ + )^{ - 1}}]^ + }{{\mathbf{s}}_i}\}  \hfill \\
  \;\;\;\; = \sum\limits_{k = 1}^S {\mathbb{E}({\mathbf{s}}_{i(k)}^ + {{\mathbf{s}}_{i(k)}}){{({{\mathbf{H}}_{ii}}{\mathbf{H}}_{ii}^ + )}^{ - 1}}_{(kk)}}  \hfill \\
  \;\;\;\; = \sum\limits_{k = 1}^S {{Q_{i(k)}}{{({{\mathbf{H}}_{ii}}{\mathbf{H}}_{ii}^ + )}^{ - 1}}_{(kk)}}  \hfill \\
  \;\;\;\; =\sum\limits_{k = 1}^S {{P_{i(k)}}}  \hfill \\
\end{gathered}, \  \tag{20}\]
where ${P_{i(k)}}$ is the transmission power transmitted by $B{S_i}$ over the $k{\text{th}}$ sub-channel, ${Q_{i(k)}}$ is the UE received power transmitted from $B{S_i}$ over the $k{\text{th}}$ sub-channel, ${({{\mathbf{H}}_{ii}}{\mathbf{H}}_{ii}^ + )^{ - 1}}_{(kk)}$ is the element located at the $k{\text{th}}$ row and the $k{\text{th}}$ column in the matrix ${({{\mathbf{H}}_{ii}}{\mathbf{H}}_{ii}^ + )^{ - 1}}$. Considering
${{P_{i(k)}} = \mathbb{E}({\mathbf{s}}_{i(k)}^ + {{\mathbf{s}}_{i(k)}}){({{\mathbf{H}}_{ii}}{\mathbf{H}}_{ii}^ + )^{ - 1}}_{(kk)}}$, the UE received power transmitted from $B{S_i}$ over the $k{\text{th}}$ sub-channel is expressed by

\[\begin{gathered}
  {Q_{i(k)}}= \mathbb{E}({\mathbf{s}}_{i(k)}^ + {{\mathbf{s}}_{i(k)}}) \hfill \\
   \;\;\;\;\;\;\;\; = \frac{{{P_{i(k)}}}}{{{{({{\mathbf{H}}_{ii}}{\mathbf{H}}_{ii}^ + )}^{ - 1}}_{(kk)}}} \hfill \\
\end{gathered}. \ \tag{21}\]

Assuming that the zero-forcing precoding is also adopted by the interfering BSs, the interference power over the $k{\text{th}}$ sub-channel in $UE{G_i}$ is obtained as
\[\begin{gathered}
  I_{i(k)}^{ZF} = \mathbb{E}[{(\sum\limits_{u \ne i,B{S_u} \in {\Pi _{BS}}} {{\mathbf{H}}_{iu}^{(k)}{{\mathbf{F}}_u}} {{\mathbf{s}}_u})^ + }(\sum\limits_{u \ne i,B{S_u} \in {\Pi _{BS}}} {{\mathbf{H}}_{iu}^{(k)}{{\mathbf{F}}_u}} {{\mathbf{s}}_u})] \hfill \\
  {\text{      }} = \sum\limits_{u \ne i,B{S_u} \in {\Pi _{BS}}} {\sum\limits_{k = 1}^S {\frac{{{P_{u(k)}}}}{{{{({{\mathbf{H}}_{uu}}{\mathbf{H}}_{uu}^ + )}^{ - 1}}_{(kk)}}}{{({\mathbf{F}}_u^ + {\mathbf{H}}_{iu}^{(k) + }{\mathbf{H}}_{iu}^{(k)}{{\mathbf{F}}_u})}_{\left( {kk} \right)}}} }  \hfill \\
  {\text{      }} = \sum\limits_{u \ne i,B{S_u} \in {\Pi _{BS}}} {\sum\limits_{k = 1}^S {\left[ \begin{gathered}
  \frac{{{P_{u(k)}}}}{{{{({{\mathbf{H}}_{uu}}{\mathbf{H}}_{uu}^ + )}^{ - 1}}_{(kk)}}} \hfill \\
   \times \frac{{\beta {w_{iu}}}}{{||{x_{B{S_u}}} - {x_{B{S_i}}} - {x_{off}}|{|^\alpha }}} \hfill \\
   \times {({\mathbf{F}}_u^ + {\mathbf{h}}_{iu}^{(k) + }{\mathbf{h}}_{iu}^{(k)}{{\mathbf{F}}_u})_{\left( {kk} \right)}} \hfill \\
\end{gathered}  \right]} }  \hfill \\
\end{gathered},\tag{22} \]
where ${\mathbf{H}}_{iu}^{(k)}$ is the $k{\text{th}}$ row of channel matrix ${{\mathbf{H}}_{iu}}$, which corresponds to the $k{\text{th}}$ sub-channel in $UE{G_i}$; ${P_{u(k)}}$ is the transmission power transmitted by $B{S_u}$ over the $k{\text{th}}$ sub-channel. ${{\mathbf{h}}_{iu}}$ is the $S \times {N_T}$ small scale fading matrix between $B{S_u}$ and $UE{G_i}$. Each element of ${{\mathbf{h}}_{iu}}$ is assumed to follow an \emph{i.i.d.} complex Gaussian distribution with zero mean and unit variance.

Based on the derivation in (16), the average interference over the $k{\text{th}}$ sub-channel in $UE{G_i}$ is derived as
\[\begin{gathered}
  \mathbb{E}(I_{i(k)}^{ZF}) \hfill \\
   = \mathbb{E}\left\{ {\sum\limits_{u \ne i,B{S_u} \in {\Pi _{BS}}} {\sum\limits_{k = 1}^S {\left[ \begin{gathered}
  \frac{{{P_{u(k)}}}}{{{{({{\mathbf{H}}_{uu}}{\mathbf{H}}_{uu}^ + )}^{ - 1}}_{(kk)}}} \hfill \\
   \times \frac{{\beta {w_{iu}}}}{{||{x_{B{S_u}}} - {x_{B{S_i}}} - {x_{off}}|{|^\alpha }}} \hfill \\
   \times {({\mathbf{F}}_u^ + {\mathbf{h}}_{iu}^{(k) + }{\mathbf{h}}_{iu}^{(k)}{{\mathbf{F}}_u})_{kk}} \hfill \\
\end{gathered}  \right]} } } \right\} \hfill \\
   = \mathbb{E}\left\{ {\sum\limits_{u \ne i,B{S_u} \in {\Pi _{BS}}} {\sum\limits_{k = 1}^S {\left\{ \begin{gathered}
  \frac{{{P_{u(k)}}}}{{{{({{\mathbf{H}}_{uu}}{\mathbf{H}}_{uu}^ + )}^{ - 1}}_{(kk)}}} \hfill \\
   \times \frac{{\beta {w_{iu}}}}{{||{x_{B{S_u}}} - {x_{B{S_i}}} - {x_{off}}|{|^\alpha }}} \hfill \\
   \times {[{\mathbf{F}}_u^ + \mathbb{E}({\mathbf{h}}_{iu}^{(k) + }{\mathbf{h}}_{iu}^{(k)}){{\mathbf{F}}_u}]_{kk}} \hfill \\
\end{gathered}  \right\}} } } \right\} \hfill \\
\end{gathered}.\tag{23} \]

Based on the result in \cite{Telatar_Capacity}, we have the following property
\[\mathbb{E}({\mathbf{h}}_{iu}^{(k) + }{\mathbf{h}}_{iu}^{(k)}) = {{\mathbf{I}}_{S \times S}}.\ \tag{24}\]

Therefore, combining (23) and (24), the average interference over the $k{\text{th}}$ sub-channel in $UE{G_i}$ is expressed by
\[\begin{gathered}
  \mathbb{E}(I_{i(k)}^{ZF}) \hfill \\
   = \mathbb{E}\left\{ {\sum\limits_{u \ne i,B{S_u} \in {\Pi _{BS}}} {\sum\limits_{k = 1}^S {\left[ \begin{gathered}
  \frac{{{P_{u(k)}}}}{{{{({{\mathbf{H}}_{uu}}{\mathbf{H}}_{uu}^ + )}^{ - 1}}_{(kk)}}} \hfill \\
   \times \frac{{\beta {w_{iu}}}}{{||{x_{B{S_u}}} - {x_{B{S_i}}} - {x_{off}}|{|^\alpha }}} \hfill \\
   \times {({\mathbf{F}}_u^ + {{\mathbf{F}}_u})_{kk}} \hfill \\
\end{gathered}  \right]} } } \right\} \hfill \\
   = \mathbb{E}\left( {\sum\limits_{u \ne i,B{S_u} \in {\Pi _{BS}}} {\sum\limits_{k = 1}^S {{P_{u(k)}}\frac{{\beta {w_{iu}}}}{{||{x_{B{S_u}}} - {x_{B{S_i}}} - {x_{off}}|{|^\alpha }}}} } } \right) \hfill \\
   = \mathbb{E}\left( {\sum\limits_{u \ne i,B{S_u} \in {\Pi _{BS}}} {{P_u}\frac{{\beta {w_{iu}}}}{{||{x_{B{S_u}}} - {x_{B{S_i}}} - {x_{off}}|{|^\alpha }}}} } \right) \hfill \\
\end{gathered},\tag{25} \]
where ${P_u} = \sum\limits_{k = 1}^S {{P_{u(k)}}} $ is the total transmission power transmitted by $B{S_u}$.
Capitalizing on (7), (10), (16) and (25), the average interference over the $k{\text{th}}$ sub-channel in $UE{G_i}$ is derived as
\[\begin{gathered}
  I_{i(k)\_avg}^{ZF}({x_{off}}) \hfill \\
   = \frac{{\beta \mathbb{E}({w_{iu}})\mathbb{E}({P_u})}}{{{\zeta ^{(1)}}}}\int\limits_{{\mathbb{R}^2}} {\frac{1}{{||x + {x_{off}}|{|^\alpha }}}{\zeta ^{(2)}}(||x||)dx}  \hfill \\
\end{gathered}.\tag{26} \]

In this paper, every UE equipped with a single antenna is assumed to be allocated with the bandwidth ${B_W}$ for data transmission. The total UEG bandwidth used for the data transmission is thus $S \cdot {B_W}$. Considering the spatial multiplexing scheme of multi-antenna systems, the data stream transmitted by a sub-channel is extended over the total UEG bandwidth. It is assumed that the noise is negligible in this paper \cite{Ge_Capacity}. To simplify the derivation, the average interference in (26) is used to calculate the capacity over the $k{\text{th}}$ sub-channel in $UE{G_i}$. As a consequence, the capacity of the $k{\text{th}}$ sub-channel in the UEG $UE{G_i}$ is expressed as
\[{C_{i(k)}} = S \cdot {B_W}{\log _2}\left\{ 1 + \frac{{{P_{i(k)}}\frac{{\beta {w_{ii}}}}{{||{x_{off}}|{|^\alpha }}}{{[{{({{\mathbf{h}}_{ii}}{\mathbf{h}}_{ii}^ + )}^{ - 1}}_{(kk)}]}^{ - 1}}}}{{I_{i(k)\_avg}^{ZF}({{x}_{off}})}}\right\}. \ \tag{27}\]

The term of ${[{({{\mathbf{h}}_{ii}}{\mathbf{h}}_{ii}^ + )^{ - 1}}_{(kk)}]^{ - 1}}$ is the random variable which governed by a Chi-square distribution. Moreover, the PDF of ${[{({{\mathbf{h}}_{ii}}{\mathbf{h}}_{ii}^ + )^{ - 1}}_{(kk)}]^{ - 1}}$ is expressed as \cite{Wang_Scheduling}
\[{\gamma _{{{[{{({{\mathbf{h}}_{jj}}{\mathbf{h}}_{jj}^ + )}^{ - 1}}_{(kk)}]}^{ - 1}}}}(\ell ) = \frac{{{\ell^{{N_T} - S}}{e^{ - \ell }}}}{{({N_T} - S)!}},\quad\ell  \geqslant 0.\ \tag{28}\]
Furthermore, the spectrum efficiency of $UE{G_i}$ in a \emph{typical} cell is derived as
\[\begin{gathered}
  S{E_{UE{G_i}}} = \frac{{\sum\limits_{k = 1}^S {{C_{i(k)}}} }}{{S{B_W}}} \hfill \\
  {\kern 1pt} {\kern 1pt} {\kern 1pt} {\kern 1pt} {\kern 1pt} {\kern 1pt} {\kern 1pt} {\kern 1pt} {\kern 1pt} {\kern 1pt} {\kern 1pt} {\kern 1pt} {\kern 1pt} {\kern 1pt} {\kern 1pt} {\kern 1pt} {\kern 1pt} {\kern 1pt} {\kern 1pt} {\kern 1pt} {\kern 1pt} {\kern 1pt} {\kern 1pt} {\kern 1pt} {\kern 1pt} {\kern 1pt} {\kern 1pt} {\kern 1pt} {\kern 1pt} {\kern 1pt} {\kern 1pt} {\kern 1pt} {\kern 1pt} {\kern 1pt} {\kern 1pt}  = \sum\limits_{k = 1}^S {\left\{ {{{\log }_2}\left\{ {1 + \frac{\xi }{S}{{[{{({{\mathbf{h}}_{ii}}{\mathbf{h}}_{ii}^ + )}^{ - 1}}_{(kk)}]}^{ - 1}}} \right\}} \right\}}  \hfill \\
\end{gathered},\tag{29a} \]
where $\xi $ is defined as the large scale SINR environment factor over wireless channels which is given as
\[\xi  = \frac{{{P_{i(k)}}S\frac{{\beta {w_{ii}}}}{{||{{x}_{off}}|{|^\alpha }}}}}{{I_{i(k)\_avg}^{ZF}({{x}_{off}})}}.\ \tag{29b}\]

Using Jensen's inequality and the mean of Chi-square distribution \cite{Masouros_Large, Paulraj_Introduction}, an upper bound for the average spectrum efficiency of $UE{G_i}$ in a \emph{typical} cell is derived as
\[\begin{array}{l}
\mathbb{E}(S{E_{UE{G_i}}})\\
 = S\mathbb{E}{\log _2}\{1 + \frac{\xi}{S}{[{({\mathbf{h}_{ii}}\mathbf{h}_{ii}^ + )^{ - 1}}_{(kk)}]^{ - 1}}\}\\
 \le S{\log _2}\{1 + \frac{\xi}{S}\mathbb{E}{{[({\mathbf{h}_{ii}}\mathbf{h}_{ii}^ + )^{ - 1}}_{(kk)}]^{ - 1}}\}\\
 = S{\log _2}[1 + \frac{\xi}{S}({N_T} - S + 1)]
\end{array}.\ \tag{30}\]

The homogeneous PPP and its associated hard core Matern point process are both stationary and isotropic \cite{Baccelli_Stochastic}. Based on the Palm theory \cite{Stoyan_Stochastic}, this feature implies that the analytical results for a \emph{typical} multi-user multi-antenna HCPP cell can be extended to the whole multi-user multi-antenna HCPP cellular network.

\subsection{Energy Efficiency of Multi-user Multi-antenna HCPP Cellular Networks}

The energy efficiency of wireless communication systems is defined as the ratio of the throughput over the consumed transmission power \cite{Ngo_Energy}. In this paper, the wireless link transmission power is assumed to be adaptively adjusted to satisfy the wireless traffic requirement \cite{Cioffi_A}, i.e., ${C_{i(k)}} = \rho $. For a given traffic rate $\rho $ and the result in (27), the wireless link transmission power in the multi-user multi-antenna HCPP cellular network is derived by
\[{P_{i(k)}} = \frac{{||{{x}_{off}}|{|^\alpha }I_{i(k)\_avg}^{ZF}({2^{\frac{\rho }{{S{B_W}}}}} - 1)}}{{\beta {w_{ii}}{{[{{({{\mathbf{h}}_{ii}}{\mathbf{h}}_{ii}^ + )}^{ - 1}}_{(kk)}]}^{ - 1}}}}.\ \tag{31}\]

For a wireless link of cellular networks, the multi-antenna system will consume additional transmission circuit block energy ${P_{RF\_chain}}$ per antenna \cite{Cui_Energy}. The total BS power is thus composed of the transmission power and the stationary power \cite{Arnold_Power}. Therefore, the average of total BS power in multi-user multi-antenna HCPP cellular networks is derived as
\[\mathbb{E}({P_{BS}}) = {N_{link}}\left[\frac{{\mathbb{E}({P_{i(k)}})}}{\eta } + {N_T}{P_{RF\_chain}}\right] + {P_{sta}},\ \tag{32}\]
where ${N_{link}}$ is the average number of active links in multi-user multi-antenna HCPP cellular networks. Considering the BS equipped with multi-antenna and the UE equipped with single antenna, the average number of active links is configured as the average active UEs in a unit BS coverage region in HCPP cellular networks, which is calculated by the density of UEs over the density of BSs, i.e., ${N_{link}} = {{{\lambda _M}} \mathord{\left/{\vphantom {{{\lambda _M}} {{\zeta ^{(1)}}}}} \right.
 \kern-\nulldelimiterspace} {{\zeta ^{(1)}}}}$. $\mathbb{E}({P_{i(k)}})$ is the average link transmission power. $\eta $ is the average efficiency of RF circuit for a BS; ${P_{sta}}$ is the stationary power for a BS.

Without loss of generality, the total BS traffic is summed by the traffic rate over all wireless links. The average of total BS traffic is derived as
\[\mathbb{E}\left( {{T_{BS}}} \right) = {N_{link}}\mathbb{E}(\rho ) = {N_{link}}\frac{{\theta {\rho _{\min }}}}{{\theta  - 1}}.\ \tag{33}\]

Therefore, the average energy efficiency of multi-user multi-antenna HCPP cellular networks is derived as
\[EE = \frac{{\mathbb{E}\left( {{T_{BS}}} \right)}}{{\mathbb{E}\left( {{P_{BS}}} \right)}} = \frac{{\frac{{\theta {\rho _{\min }}}}{{\theta  - 1}}}}{{\left[\frac{{\mathbb{E}({P_{i(k)}})}}{\eta } + {N_T}{P_{RF\_chain}}\right] + \frac{{{\zeta ^{(1)}}}}{{{\lambda _M}}}{P_{sta}}}}.\ \tag{34}\]

\section{Numerical Results and Discussions}
\label{sec5}
Based on the proposed spectrum efficiency and energy efficiency models in Sections IV and V, numerical results are analyzed in detail. Moreover, analysis results are validated through MC simulations in multi-user multi-antenna HCPP cellular networks. The default parameters used for the simulations are as follows \cite{Xiang_Energy,Cui_Energy,Arnold_Power}: ${\lambda _P} = 1/(\pi *{800^2})$, $\delta  = 500$ meters, $\beta  =  - 31.54$ dB, $\alpha  = 3.8$, ${\sigma _s} = 6$, $\mathbb{E}({P_{u(k)}}) = 2$ W, $\theta  = 1.8$, $\eta  = 0.38$, ${P_{RF\_chain}} = 50$ milliwatt (mW), ${P_{sta}} = 45.5$ W, ${N_{link}} = 30$. The maximum BS transmission power is assumed to be 40 W over 200 KHz carrier bandwidth \cite{Xiang_Energy}. Furthermore, the maximum transmission power over a wireless link with 10 KHz carrier bandwidth is configured as 2 W in the multi-user multi-antenna HCPP cellular networks \cite{Cui_Energy,Arnold_Power}. We assume that the wireless link is interrupted if the corresponding transmission power is larger than 2 W.

\begin{figure}[H]
\vspace{0.1in}
\centerline{\includegraphics[width=9.5cm,draft=false]{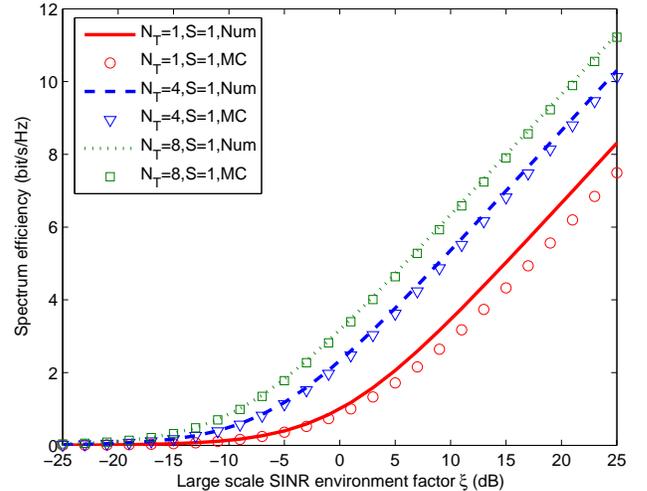}}
\caption{\small Impact of the large scale SINR environment factor on the spectrum efficiency of multi-user multi-antenna HCPP cellular networks considering different number of transmit antennas at the BS.}
\end{figure}

Fig. 6 illustrates the impact of the large scale SINR environment factor $\xi $ on the spectrum efficiency of multi-user multi-antenna HCPP cellular networks for different numbers of transmission antennas at the BS. In this case, the number of antennas at the UEG is $S = 1$. We observe that when the number of transmit antennas is fixed, the spectrum efficiency of multi-user multi-antenna HCPP cellular networks increases with increasing $\xi $. When $\xi $ is fixed, the spectrum efficiency of multi-user multi-antenna HCPP cellular networks increases with increasing the number of transmit antennas at the BS. The MC simulation curves agree well with the numerical curves in Fig. 6. Since numerical curves are plotted by the upper bound of the spectrum efficiency, MC simulation values are always less than or equal to numerical values in Fig. 6.

\begin{figure}
\vspace{0.1in}
\centerline{\includegraphics[width=9.5cm,draft=false]{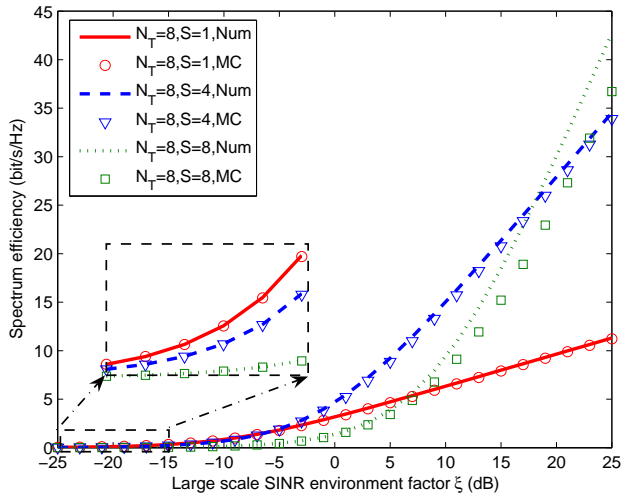}}
\caption{\small Impact of the large scale SINR environment factor on the spectrum efficiency of multi-user multi-antenna HCPP cellular networks considering different number of receive antennas at the UEG.}
\end{figure}

Fig. 7 analyzes the impact of $\xi $ on the spectrum efficiency of multi-user multi-antenna HCPP cellular networks considering different receive antenna numbers in the UEG, for the case where the number of transmit antennas at the BS equals 8. We observe that when number of receive antennas at the UEG is fixed, the spectrum efficiency of multi-user multi-antenna HCPP cellular networks increases with the increasing of $\xi $. When $\xi $ is large, the spectrum efficiency of multi-user multi-antenna HCPP cellular networks increases with increasing the number of receive antennas at the UEG. When the large scale SINR environment factor $\xi $ is low, the spectrum efficiency of multi-user multi-antenna HCPP cellular networks increases with decreasing the number of receive antennas at the UEG. The MC simulation curves exhibit a good match with the numerical curves in Fig. 7.

\begin{figure}
\vspace{0.1in}
\centerline{\includegraphics[width=9.5cm,draft=false]{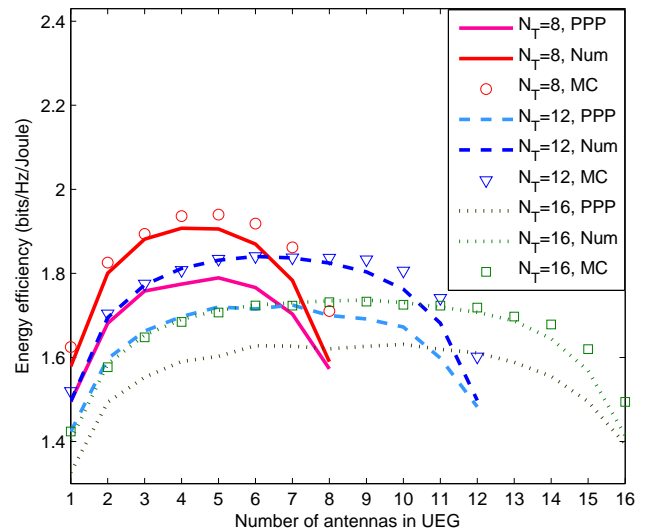}}
\caption{\small Energy efficiency of multi-user multi-antenna HCPP and PPP cellular networks with respect to the number of antennas at the UEG and the BS.}
\end{figure}

Fig. 8 illustrates the energy efficiency of multi-user multi-antenna HCPP and PPP cellular networks with respect to the number of antennas at the UEG and the BS. When the number of antennas at the BS is fixed, both numerical and MC simulation results illustrate that the energy efficiency of multi-user multi-antenna HCPP and PPP cellular networks first increases with increasing the number of antennas at the UEG. When the number of antennas at the UEG is larger than the threshold which corresponds the maximal value of the energy efficiency in this curve, both numerical and MC simulation results show that the energy efficiency of cellular networks decrease with increasing the number of antennas at the UEG. There exist different maximal energy efficiency values of multi-user multi-antenna HCPP cellular networks when BSs are integrated with different antenna numbers. In numerical results, the maximal energy efficiency values are 1.9, 1.84 and 1.72 bits/Hz/Joule, which corresponds to the number of antennas at the BS as 8, 12 and 16, respectively. Moreover, both numerical and MC simulation results indicate that the available maximal energy efficiency values of multi-user multi-antenna HCPP cellular networks decreases with increasing the number of antennas at the BS. Meanwhile, the energy efficiency of PPP cellular networks is less than the energy efficiency of HCPP cellular networks. This result is also validated in Fig. 9-11.

\begin{figure}
\vspace{0.1in}
\centerline{\includegraphics[width=9.5cm,draft=false]{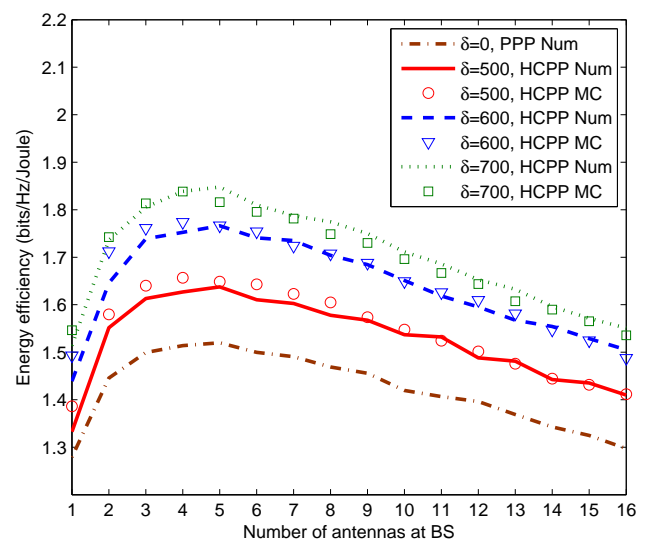}}
\caption{\small Energy efficiency of multi-user multi-antenna HCPP and PPP cellular networks with respect to the number of antennas at the BS and the minimum distance $\delta$ in adjacent BSs.}
\end{figure}

Without loss of generality, the number of antennas at the BSs and the UEGs in the multi-user multi-antenna HCPP and PPP cellular networks are configured as equal in Fig. 9-11. Fig. 9 shows the energy efficiency of multi-user multi-antenna HCPP and PPP cellular networks with respect to the number of antennas at BS and the minimum distance $\delta $. When the minimum distance $\delta $ is fixed, both numerical and MC simulation results illustrate that the energy efficiency of multi-user multi-antenna HCPP and PPP cellular networks first increases with increasing the number of antennas at the BS. When the number of antennas at the BS is larger than the threshold, both numerical and MC simulation results show that the energy efficiency of multi-user multi-antenna HCPP and PPP cellular networks decreases with increasing the number of antennas at the BS. This result is different with the energy efficiency respect to the number of antennas at the BS in massive MIMO systems \cite{Ngo_Energy}. One of reasons is that the small scale fading effect is ignored for wireless channels in massive MIMO systems \cite{Marzetta}. On the contrary, the small scale fading effect is considered for the capacity and the interference modeling in this paper. We observe that there exist different maximal energy efficiency values of multi-user multi-antenna HCPP cellular networks under different minimum distances. The maximal energy efficiency values are 1.85, 1.73 and 1.63 bits/Hz/Joule, corresponding to the minimum distances of 300, 400 and 500 meter, respectively. When the number of antennas at the BS is fixed, both numerical and MC simulation results consistently demonstrate that the available maximal energy efficiency of multi-user multi-antenna HCPP cellular networks increases with increasing the minimum distance $\delta $. Based on our previous results in \cite{Humar11,Xiang_Energy}, there exist an optimal value of cell size, e.g. 1200 meters, corresponding to the maximal energy efficiency of cellular networks when the stationary power ${P_{sta}}$, i.e., the embodied power, is considered for the BS power consumption. When the minimum distance $\delta $ is less than the optimal value of cell size, the maximal energy efficiency of HCPP cellular networks increases with increasing the minimum distance.


\begin{figure}
\vspace{0.1in}
\centerline{\includegraphics[width=9.5cm,draft=false]{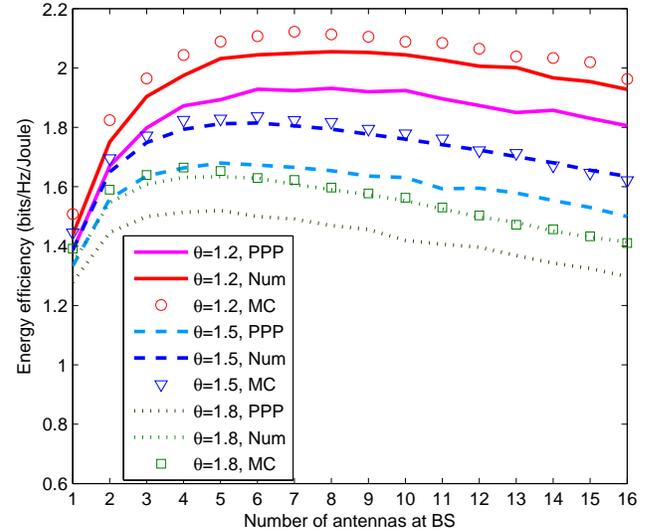}}
\caption{\small Energy efficiency of multi-user multi-antenna HCPP and PPP cellular networks with respect to the number of antennas at the BS and the traffic heaviness index $\theta$.}
\end{figure}

Fig. 10 analyzes the energy efficiency of multi-user multi-antenna HCPP and PPP cellular networks with respect to the number of antennas at the BS and the traffic heaviness index $\theta $, where the minimum traffic rate over the unit bandwidth is fixed as ${{{\rho _{\min }}} \mathord{\left/
 {\vphantom {{{\rho _{\min }}} {{B_W}}}} \right.
 \kern-\nulldelimiterspace} {{B_W}}} = 2$. When the number of antennas at the BS is fixed, both numerical and MC simulation results illustrate that the energy efficiency of multi-user multi-antenna HCPP and PPP cellular networks increases with decreasing the traffic heaviness index $\theta $. There exist different maximal energy efficiency values of multi-user multi-antenna HCPP cellular networks under different traffic heaviness indices. The maximal energy efficiency values are 2.06, 1.81 and 1.64 bits/Hz/Joule, which corresponds to the traffic heaviness index as 1.2, 1.5 and 1.8, respectively. When the number of antennas at the BS is fixed, both numerical and MC simulation results demonstrate that available maximal energy efficiency of multi-user multi-antenna HCPP and PPP cellular networks decreases with increasing the traffic heaviness index $\theta $. The increasing of the traffic heaviness index implies that the burst of traffic is increased. The increased burst of traffic will decrease the utilization efficiency of the wireless channel capacity. As a result, the available maximal energy efficiency is decreased with increasing the traffic burst in HCPP cellular networks.

\begin{figure}
\vspace{0.1in}
\centerline{\includegraphics[width=9.5cm,draft=false]{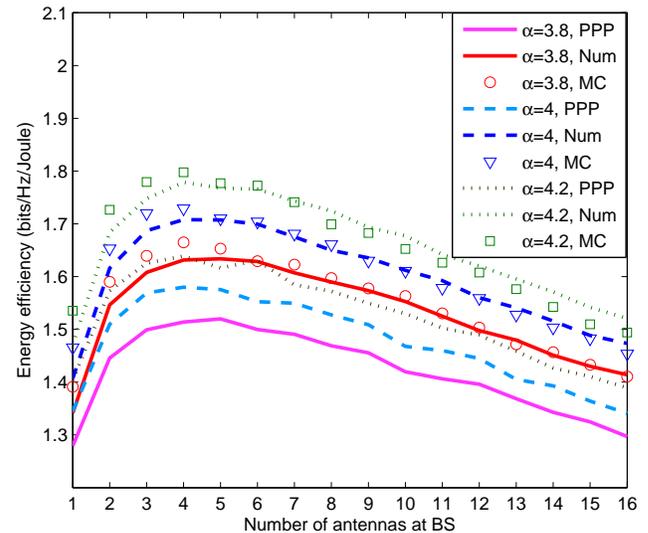}}
\caption{\small Energy efficiency of multi-user multi-antenna HCPP and PPP cellular networks with respect to the number of antennas at the BS and the path loss coefficient $\alpha$.}
\end{figure}

Finally, the impact of path loss coefficient on the energy efficiency of multi-user multi-antenna HCPP and PPP cellular networks is evaluated in Fig. 11. When the number of antennas at the BS is fixed, both numerical and MC simulation results show that the energy efficiency of multi-user multi-antenna HCPP and PPP cellular networks increases with increasing of the path loss coefficient $\alpha $. Moreover, the maximum energy efficiency with three different path loss coefficients are 1.78, 1.71 and 1.63 bits/Hz/Joule, which correspond to the path loss coefficient as 4.2, 4.0 and 3.8, respectively. The interference fading becomes severer when the path loss coefficient is increased, which leads to the higher spectrum efficiency in wireless channels. As a consequence, the available maximal energy efficiency is increased with increasing the spectrum efficiency in HCPP cellular networks.

\section{Conclusions}
\label{sec6}

We proposed an energy efficiency assessment for multi-user multi-antenna HCPP cellular networks considering the minimum distance constraint in adjacent BSs. This assessment was obtained by considering an average interference model for multi-antenna HCPP cellular networks with the shadowing and small scale fading over wireless channels. Based on the zero-forcing precoding method, a spectrum efficiency assessment was also obtained for multi-user multi-antenna HCPP cellular networks. Based on the proposed energy efficiency model of multi-user multi-antenna HCPP cellular networks, numerical results have shown that there exists the maximal energy efficiency in multi-user multi-antenna HCPP cellular networks. Our analysis indicates that the maximal energy efficiency of multi-user multi-antenna HCPP cellular networks decreases with increasing the number of transmit antennas at the BSs. Moreover, the maximal energy efficiency of multi-user multi-antenna HCPP cellular networks was shown to depend on the wireless traffic distribution, the wireless channel and the minimum distance in adjacent BSs. Furthermore, the comparison between HCPP and PPP cellular networks implies that the energy efficiency of the conventional PPP cellular networks is underestimated when the minimum distance in adjacent BSs is ignored. Interesting topics for future work include the investigation of the energy efficiency of random cellular networks under massive MIMO scenarios and the UE association based on channel conditions.

\end{document}